\documentclass[aps,pre,twocolumn,longbibliography,notitlepage,floatfix,nofootinbib]{revtex4-1}

\usepackage{bm,amsmath,amssymb,graphicx,stmaryrd,float,subcaption,upgreek,sidecap,MnSymbol,mathtools}
\usepackage{epstopdf}
\usepackage[toc]{appendix}
\usepackage[justification=raggedright]{caption}
\usepackage[abs]{overpic}
\newcommand{\uvc}[1]{\bm{\mathrm{\hat #1}}} 
\usepackage{comment,footnote}
\usepackage[thinc]{esdiff}
\usepackage{tikz}

\usepackage[format=plain,labelfont={bf,small},textfont=small,justification=raggedright,singlelinecheck=false]{caption}
\usepackage{hyperref}

\usepackage{fullpage}

\begin{document}
	\title{Cutting holes in bistable folds}
	\author{T. Yu}
	\email{tiany@princeton.edu}
	\affiliation{Department of Civil and Environmental Engineering, \\Princeton University, Princeton, NJ 08544, USA}
	
	\author{I. Andrade-Silva}
	\email{ignacio.andrade-silva@univ-amu.fr}
	\affiliation{Aix-Marseille University, CNRS, IUSTI, Marseille 13013, France}
		
	\author{M. A. Dias}
	\email{marcelo.dias@ed.ac.uk}
	\affiliation{Institute for Infrastructure \& Environment, School of Engineering, The University of Edinburgh, Edinburgh EH9 3FG, UK \\
	Department of Mechanical and Production Engineering, Aarhus University, 8000 Aarhus C}
	
	\author{J. A. Hanna}
	\email{jhanna@unr.edu}
	\affiliation{Department of Mechanical Engineering, University of Nevada, 1664 N. Virginia St.\ (0312),  Reno, NV 89557-0312, USA}
	\date{\today}
\begin{abstract}
A folded disk is bistable, as it can be popped through to an inverted state with elastic energy localized in a small, highly-deformed region on the fold. Cutting out this singularity relaxes the surrounding material and leads to a loss of bistability when the hole dimensions reach a critical size.  These dimensions are strongly anisotropic and feature a surprising re-entrant behavior, such that removal of additional material can re-stabilize the inverted state.  A model of the surface as a wide annular developable strip is found to capture the qualitative observations in experiments and simulations.  These phenomena are consequential to the mechanics and design of crumpled elastic sheets, developable surfaces, origami and kirigami, and other deployable and compliant structures.
\end{abstract}
\maketitle

The role of elastic singularities in the deformation of thin sheets and shells is still poorly understood, despite a quarter of a century of intense investigation into their geometry and energetics \cite{AmirbayatHearle86-1, AmirbayatHearle86-2, BenAmarPomeau97, Chaieb98, CerdaMahadevan98, MoraBoudaoud02, lobkovsky1995scaling, DiDonna02, LiangWitten05, FarmerCalladine05, Nasto13, ChopinKudrolli16, Yang18, Moshe19, Elder19}.
Over the years, several perspectives have emerged, viewing these localized high-energy regions as a manifestation of spontaneous condensation of both curvature and stretching  
\cite{Witten07, Das07, Schroll11, Mellado11},
sources of rigidity \cite{Balkcom09, Tallinen10, Guven13dipoles}, 
or an organizing framework for random crumpling \cite{CambouMenon11},
regular patterns \cite{Thompson15, Timounay20},
or dynamics \cite{Hamm04, walsh2011weakening}.
Our interest in the current study was driven by Witten's observation \cite{witten09spontaneous} that the excision of such regions of focused elastic energy leads to significant relaxation of neighboring regions of material, and the indication that such surgery should also modify the rigidity and stability landscape of any surrounding structure.
The proximity of the edge of regression, or other virtual singularities living outside nominally inextensible surfaces, has been qualitatively linked to the structural stiffness response \cite{korte10triangular, dias2012geometric, Couturier13}.  
Another thread in this work is the question of multistability of systems of creases and facets with competing flexibilities, including origamic analogs of elastic singularities \cite{Seffen12, seffen18spherical, hanna2014waterbomb, silverberg2014using, Waitukaitis15, lechenault15generic, andrade2019foldable, gillman2018truss}. 

Our model system is a single fold in an elastic disk, and the singular structure formed by popping it through with a thumb.  
This is perhaps the simplest bistable ``foldable cone'' examined in \cite{lechenault15generic}.  
In \cite{walker18shape} it was noted that a small hole reduced the energy barrier to pop through a fold to its inverted state, but this was not pursued to its logical conclusion, the complete elimination of the barrier with a sufficiently large hole.
In this note, we employ theory, numerics, and experiment to capture the complex behavior of a fold after removal of its singularity and a variable zone of surrounding material.
We find that cutting a hole of sufficient size around the singularity leads to a loss of bistability through a fold bifurcation that destabilizes the inverted state.  There is significant anisotropy in the critical hole dimensions, such that a narrow slit aligned along the crease can be as large as the disk without destroying bistability.  We also observe a curious re-entrant behavior of the stability diagram for small elliptical holes aligned perpendicular to the crease, which can in some parameter ranges more effectively eliminate bistable behavior than a larger circular hole. 
We demonstrate the surprising applicability of a developable ribbon model to this class of wide, topologically annular shapes.

\begin{figure*}[h!]
	\centering
	\captionsetup[subfigure]{labelfont=normalfont,textfont=normalfont}
	\centering
	\includegraphics[width=0.9\textwidth]{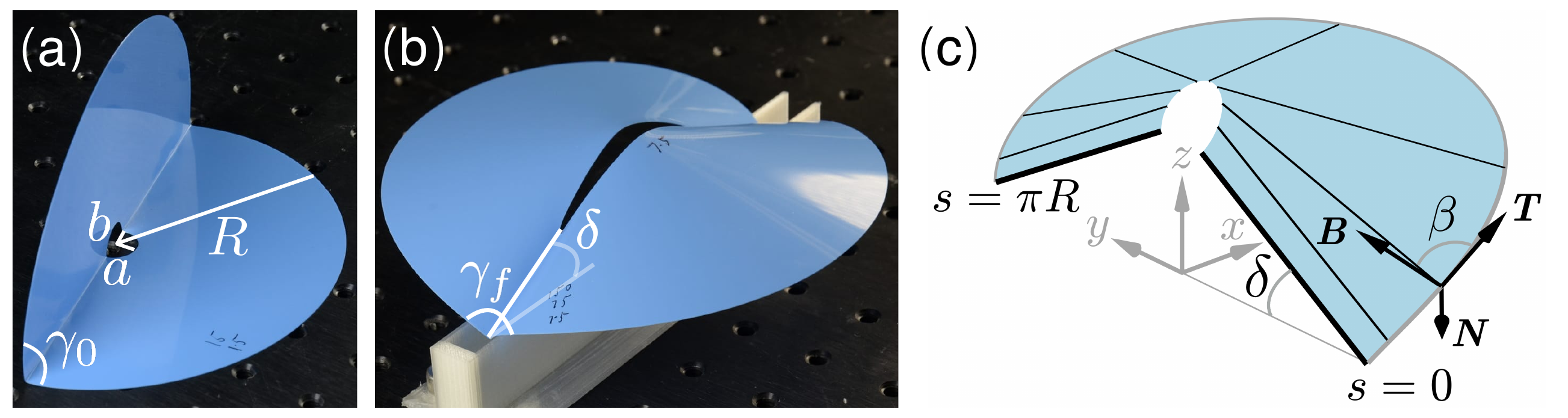}
	\caption{Photographs of (a) folded and (b) inverted states of disks with different hole dimensions, showing the disk radius $R$, hole semiaxes $a$ and $b$ perpendicular and parallel to the crease, rest and final crease angles $\gamma_0$ and $\gamma_f$, and an angle $\delta$ characterizing the inverted state.
(c) Rendering of one half of an inverted state in a developable strip model, bounded by the crease.  The directrix is the outer circumference, parameterized by $0 \le s \le \pi R$ for one half of the disk, and carrying a Darboux frame $(\bm{T},\bm{N},\bm{B})$.   The generators make a local angle $\beta$ with the tangent.}\label{fig:foldsparameter}
\end{figure*}

Experiments were performed on disks of radius $1 \le R \le 10$ cm, thickness $t=$ 0.005 in (0.127 mm), 0.003 in (0.076 mm), or 0.002 in (0.051 mm), and prescribed central elliptical hole geometries (semiaxes $a$ and $b$ perpendicular and parallel to the eventual crease), obtained by cutting (Cameo 3, Silhouette America, Lindon, UT) polyester shim stock (Artus Corp., Englewood, NJ) and subsequently creasing along a diameter using a vise set to a prescribed position and held for two seconds.
The structure was then flexed by inverting it once or twice and allowed to relax for five minutes before bistability tests were performed, with the sample hanging such that the crease was vertical to minimize gravitational effects on the bistability of the thinnest sheets.
 We refer to the simply creased state as the \emph{folded configuration} 
  and the stable popped-through state as the \emph{inverted configuration}. 
 When both stable states exist, they are separated by an energy barrier corresponding to another, unstable, equilibrium state.  This barrier was examined in \cite{Walker20} for a reduced model of the surface with a single hole size. 
The folded state is characterized by a rest crease angle $\gamma_0$ over which we have little control; thickness, disk size, and hole size all contribute significantly to this value, which we report as a range spanning multiple hole sizes for a given sample set (measured from photographs of the disks).
The inverted state is characterized by a final crease angle $\gamma_f$, which is observed to be a function of radial position in experiments and numerics, and an angle $\delta$ between a side of the crease and a line connecting the ends of the two sides.  Figure \ref{fig:foldsparameter}(a-b) illustrates examples of the two states and associated parameters for two different hole geometries.
The inverted configuration reflects a competition between the unknown and uncontrollable stiffness of the crease and the bending resistance of the facets comprising the remainder of the disk, such that the overall disk size $R$ is a relevant scale that we can understand using the related concepts of ``origami length'' \cite{lechenault2014mechanical} or ``hinge index'' \cite{francis2013origami}.
We should expect an asymptotic approach to a linear scaling of critical bistable hole size with the disk radius as the latter grows and the crease becomes effectively rigid.  However, the interference of gravity also becomes more important with increasing disk radius, setting a practical limit for the experiments.

We employ an annular wide-strip model that treats the punctured disk as a developable surface outside of the crease, an approach that is reasonable for static configurations of sufficiently thin elastic sheets \cite{starostin2007shape, korte10triangular, starostin2015equilibrium, ChopinKudrolli16, dias2012geometric, MooreHealey18, yu2019bifurcations, badger2019normalized}.
In this model, we treat the crease as a generator (zero-curvature direction) with uniform final angle $\gamma_f$, an approximation we will revisit shortly below.
Full details of the model, including boundary conditions and numerical implementation, can be found in Appendix \ref{app:annulusmodel}; we sketch the important aspects here.
 The directrix $\bm{r}(s)$ forms the outer circumference of one half of the symmetric disk, parameterized by arc length $0 \le s \le \pi R$, and carries an orthonormal material (Darboux) frame of curve tangent $\bm{T} = \bm{r}'$, where a prime denotes an $s$-derivative, surface normal $\bm{N}$, and surface tangent normal $\bm{B} = \bm{T}\times\bm{N}$.
 The evolution of the frame is given by $\bm{T}'= \kappa_n \bm{N}  - \kappa_g \bm{B}$, $\bm{N}'= -\kappa_n \bm{T} + \tau_g \bm{B}$, $\bm{B}'= \kappa_g \bm{T} - \tau_g \bm{N}$, where  $\kappa_n$, $\kappa_g = -1/R$, and $\tau_g$ are the normal curvature, geodesic curvature, and geodesic torsion.
  As shown in Figure \ref{fig:foldsparameter}(c), generators lie in the $\bm{B}+\eta \bm{T}$ direction, making a local angle $\beta$ with the directrix; $\eta = \cot \beta = \tau_g / \kappa_n$.
 The shape is symmetric and given by the embedding
\begin{align}\label{eq:developabledescription} 
	\bm{X}(s,v)=\bm{r}(s)+v[\bm{B}(s)+\eta(s)\bm{T}(s)] \, , 
\end{align}
with $0 \le v \le V(s,\eta \, ; R,a,b)$ the coordinate along the generator of implicitly treated length $V \sqrt{1+\eta ^2}$.
This surface has mean curvature $H= \tfrac{\kappa_n (1+\eta^2)}{{2[1+v[\eta'+\kappa_g(1+\eta^2)]]}}$ and  area element $dA=[1+v(\eta'+\kappa_g (1+\eta^2))] \, dsdv$. 
Defining a crease stiffness per unit length $K_c$ and a facet bending rigidity $D=E t^3 /[12 (1 -\nu ^2)]$ incorporating the Young's modulus $E$ and Poisson's ratio $\nu$, the total elastic energy $U$ of a creased punctured disk can be written as 
an augmented Wunderlich functional \cite{todres2015translation, starostin2007shape, dias2012geometric, dias2015wunderlich}
with contributions from both crease and facets,

\begin{widetext}
\begin{align}
\frac{U}{2D}&= \frac{K_{c}}{D}(R-b)  \int_{\gamma_0}^{\gamma_f} \sin(\tilde\gamma_f-\gamma_0) d \tilde\gamma_f \nonumber 
  + \frac{1}{2}\int_0^{\pi R} \! \int_{0}^{V}{(2H)}^2 \,dA  \,,  \nonumber  \\
&= \frac{K_{c} R}{D} \left(1-\frac{b}{R}\right) \left[1-\cos (\gamma_f - \gamma_0)\right] + \int_0^{\pi R} YW ds  \,, \label{eq:totalenergy}
\end{align}
\end{widetext}
with $Y=\frac{\kappa^2_n (1+\eta^2)^2}{2[\eta'+\kappa_g(1+\eta^2)]} $ and $W=\ln[1+V (\eta'+\kappa_g (1+\eta^2))]$.
The crease stiffness diverges as the origami length $D/K_c$ approaches zero with the thickness \cite{lechenault2014mechanical}.
The crease contribution to the energy \eqref{eq:totalenergy} only enters the problem through the boundary conditions, and does not appear in the Euler-Lagrange equations, given by \cite{starostin2007shape,dias2015wunderlich}
\begin{align}
\bm{F}' &= \bm{0} \,, \label{eq:stripgovern1} \\
\bm{M}' + \bm{T} \times \bm{F} &=\bm{0} \,, \label{eq:stripgovern2} \\
\partial_{\kappa_n} (YW) -\eta M_1 -M_3 &=0 \,, \label{eq:stripgovern3} \\ 
\partial_\eta (YW) -(\partial_{\eta'} (YW))'-\kappa_n M_1 &=0 \label{eq:stripgovern4}\,,
\end{align}
where forces and moments, normalized by $D$, have components in the moving frame specified by $\bm{F}=F_1 \bm{T}+F_2 \bm{N}+F_3 \bm{B}$ and  $\bm{M}=M_1 \bm{T}+M_2 \bm{N}+M_3 \bm{B}$. 
Equations (\ref{eq:stripgovern1}-\ref{eq:stripgovern4}) for one half of the symmetric structure, along with an Euler angle description of the moving frame, and boundary conditions imposed at the crease, are solved using the continuation package AUTO 07P \cite{doedel2007auto}.

To explore beyond the limitations of the developable model, as well as to allow independent control of rest crease angle, crease stiffness, and material thickness not possible in experiments, simulations were performed using the commercial finite element (FE) software COMSOL Multiphysics 5.4, employing quadratic shell elements, linear-elastic material, and creases introduced using a through-thickness thermal gradient \cite{andrade2019foldable}. 
These simulations also avoid self-contact effects that are an issue with experiments at small hole sizes.  
Full simulation details can be found in Appendix \ref{app:FEA}. 

\begin{figure*}[h!]
	\centering
	\captionsetup[subfigure]{labelfont=normalfont,textfont=normalfont}
	\centering
	\includegraphics[width=0.9\textwidth]{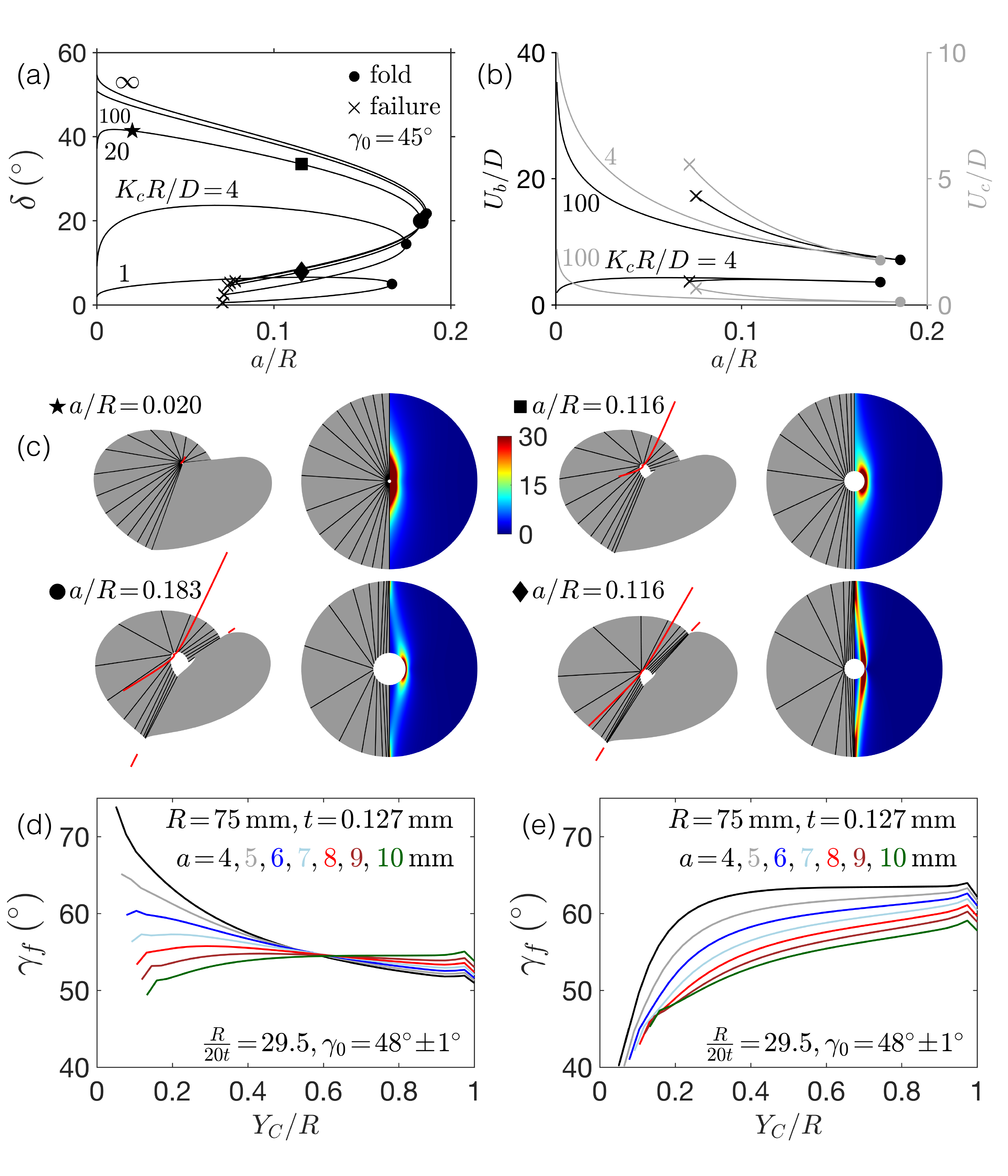}
	\caption{(a) Developable strip solutions for circular holes of radius $a$, rest crease angle $\gamma_0=45^{\circ}$, and several values of dimensionless crease stiffness $K_cR/D$.  The $K_cR/D = \infty$ curve is generated using a fixed (rigid) crease angle.  A stable and an unstable branch are created at the fold bifurcation, and the curves truncate where the developable assumption fails.  (b) Normalized bending energy $U_b/D$ and normalized crease energy $U_c/D$.  (c) Renderings of stable, near-fold-point, and unstable states for the developable model with $\gamma_0=45^{\circ}$ and $K_cR/D=20$ showing generators (black lines), edges of regression (red curves), and color maps of twice the squared mean curvature $2H^2$ with $R$ set to unity.  Generators near the crease align with it at $a/R \approx 0.116$. (d) Opening angle $\gamma_f$ along the crease coordinate $Y_C$ in finite element simulations of the stable inverted state for several values of hole radius, an outer disk radius of 75 mm, a thickness of 0.127 mm, and a crease angle of $48^{\circ} \pm 1^{\circ}$. The effective crease stiffness is $R/20t=29.5$. (e) Same for the unstable energy barrier state.  Bistability is lost between $a=10$ and $a=11$ mm.  The nonsmoothness of the curves near the boundaries is likely due to the finite mesh size, relatively large local deviations in the rest angle at the boundaries, and the possible emergence of a secondary curvature along the crease \cite{lobkovsky1995scaling}.
}\label{fig:circularsolu}
\end{figure*}

Results from the developable model (\ref{eq:stripgovern1}-\ref{eq:stripgovern4}) are shown in 
Figure \ref{fig:circularsolu}(a-c) for circular holes ($a=b$), rest crease angle $\gamma_0=45^{\circ}$, and several values of crease stiffness $K_cR/D$.   
Certain features are shared by all the solution curves in parameter space.  The stable inverted state and another unstable state are lost through a fold bifurcation at a critical hole size, which shows only moderate variation with crease stiffness.  At the end of the unstable lower branch, the curves terminate due to a failure of the embedding of the developable surface as the edge of regression approaches the sheet boundary.
For stiffer creases, a greater proportion of the energy $U$ is associated with facet bending energy $U_b$ rather than the opening of the crease.
At small hole sizes, the crease energy increases rapidly, and the facet energy does likewise for stiff creases but decreases for softer creases.
Renderings of stable, near-fold-point, and unstable inverted states of the developable model with $\gamma_0=45^{\circ}$ and $K_cR/D=20$ show the generators (black lines), edges of regression (red curves), and color maps of twice the squared mean curvature $2H^2$ (with $R$ set to unity) that appears in the facet energy.  Further details of these solutions can be found in Appendix \ref{se:ktauspecialpoints}.
A small hole size leads to a shape similar to a generalized cone, which would have a single point ``inside'' the annulus as its edge of regression, rather than the cusped curve of the more general structure.  As the hole size increases, the generators near the crease first align with it, becoming ``cylindrical'' at $a/R \approx 0.116$, echoing the qualitative features of the ridge relaxation observed in Witten's experiment \cite{witten09spontaneous}.  Past this point, the generators near the crease converge towards the outer perimeter of the annulus, so that a portion of the edge of regression lies ``outside'' it.
On a developable strip, the bending moment is inversely proportional to the distance of a point on a generator to the edge of regression.  Thus, we would expect that if the developable constraint were relaxed to allow a non-uniform crease angle $\gamma_f$, 
the inner part of the crease would open to a wider angle than the outer part for small holes and \emph{vice versa} for large holes.  Indeed we observe this effect in experimental samples.  Results on this effect from a more quantitative analysis using FE are shown in 
Figure \ref{fig:circularsolu}(d-e) for the stable inverted configuration and the unstable state on the energy barrier, respectively, for several circular hole sizes up to a value close to the loss of bistability.  A representative set of crease parameters are used (the value $R/20t$ can be considered an effective crease stiffness for the FE results, as discussed in Appendix \ref{app:FEA}). 
In the inverted configuration, the crease angle variation along the normalized distance on the crease $Y_C/R$ from the center of the hole shows the expected transition between inner and outer crease opening as the hole size increases.  The developable assumption of constant crease angle works better for large hole sizes.  In contrast, the unstable equilibrium is always more open towards the outer edge of the annulus.  Curiously, the crease angle near the hole in the unstable equilibrium is actually smaller than the rest angle.

\begin{figure*}[h!]
	\centering
	\includegraphics[width=0.9\textwidth]{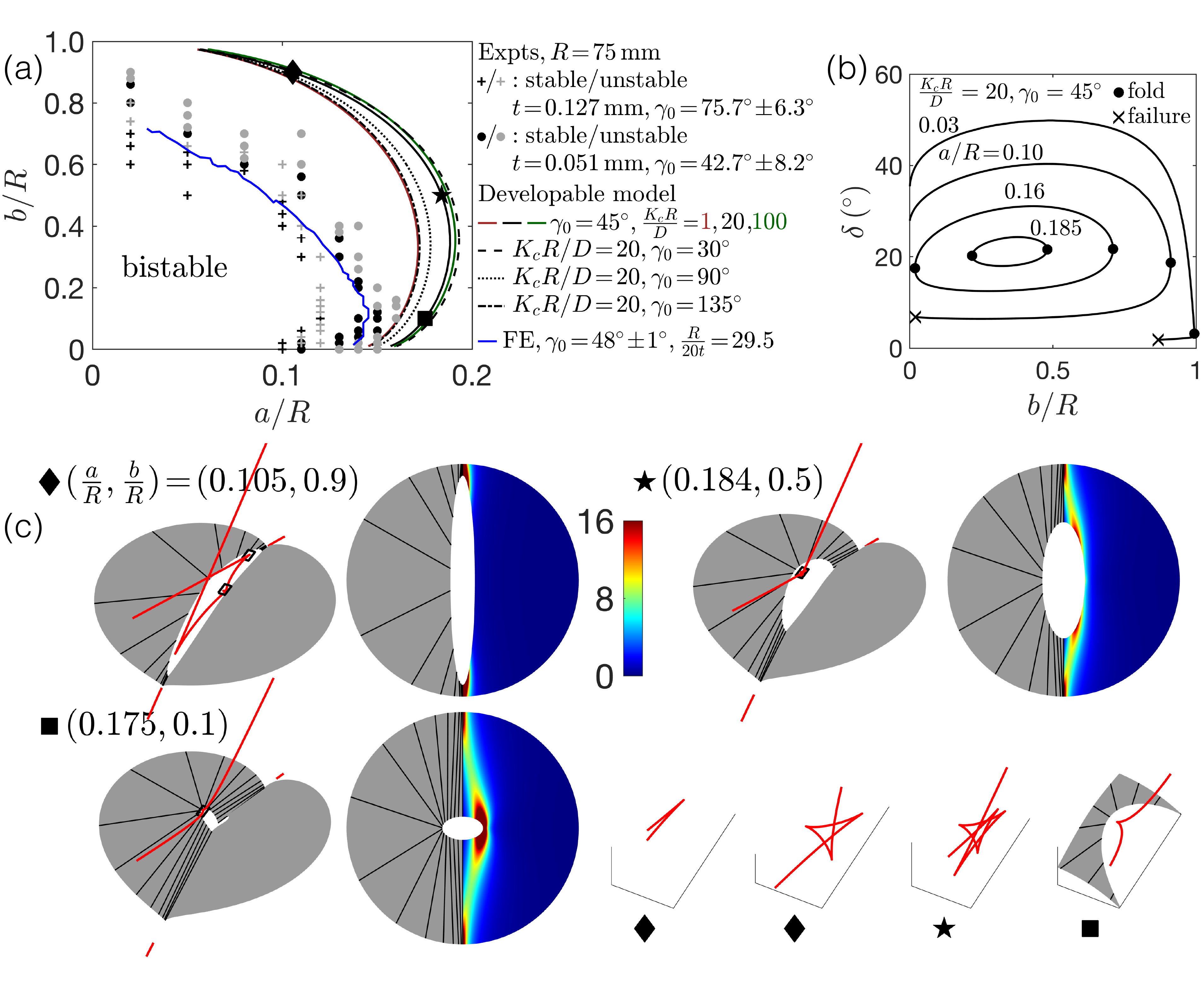}
	\caption{(a) Bistability boundaries for elliptical holes with semiaxes $a$ perpendicular to and $b$ parallel to the crease, for the developable model for several values of rest crease angle $\gamma_0$ and crease stiffness $K_cR/D$, experiments with two thicknesses (crease angles reported are average and standard deviation of $14-15$ samples at each thickness across a range of hole sizes), and FE simulations for one choice of crease parameters (using $R=75$mm and $t=0.127$mm, as in one set of experiments). (b) Solution curves, bounded by fold bifurcations or points of failure of the developable assumption, and (c) renderings of stable inverted states near the bistability boundary for the developable model with $\gamma_0=45^{\circ}$ and $K_cR/D=20$ showing generators (black lines), edges of regression (red curves), and color maps of twice the squared mean curvature $2H^2$ with $R$ set to unity.	 Close-ups of the edges of regression show complex forms.}\label{fig:phaseeps12}
\end{figure*}

We next explore the anisotropic interaction between material removal and the elasticity of the structure by considering elliptical holes with semiaxes $a$ perpendicular to and $b$ parallel to the crease.
Bistability boundaries for the developable model are shown in Figure \ref{fig:phaseeps12}(a) for several values of rest crease angle $\gamma_0$ and crease stiffness $K_cR/D$.
Also shown are boundaries determined from experiments 
on two thicknesses of material and from FE simulations of one thickness and set of crease parameters. A quantitative comparison across the three approaches is not possible. In experiments, the crease stiffness is unknown and the rest angle has a strong dependence on thickness and hole geometry; we report an average and standard deviation across a range of hole sizes in order to treat hole size as if it were an independent parameter.
Thickness appears in the stiffness in the developable model; a previous observation indicates that the origami length $D/K_c \approx 200t$ for mylar (a polyester) sheets \cite{lechenault2014mechanical}, 
although we note that the geometry of the current problem is significantly different than that of the cited work.  The crease becomes effectively rigid as the thickness vanishes.
The FE simulations use an angle close to one of the developable solutions, with an effective stiffness falling within a range in which the dependence of the developable model on stiffness is very weak.
However, the qualitative behavior of all the boundary curves is the same, and the quantitative spread is fairly small within a wide range of reasonable parameters.  Boundaries are shifted to larger hole sizes by a smaller crease angle (sharper fold) or by a stiffer crease; the experiments on thinner materials behave accordingly as having both sharper folds and stiffer creases. 
An unexpected re-entrant behavior of the boundary curves is present in all three approaches: theory, experiments, and simulations.  This means that in some parameter ranges, removing more material actually leads to a reappearance of bistable behavior.  Fixing the length of the hole axis perpendicular to the crease, elliptical holes with long axis perpendicular or parallel to the crease can be monostable while less eccentric holes are bistable.  It is also apparent that long, slit-like holes along the crease do not eliminate bistability, while small perpendicular slits do.
Solution curves, and renderings of stable inverted states near the bistability boundary, for the developable model with $\gamma_0=45^{\circ}$ and $K_cR/D=20$, are shown in Figure \ref{fig:phaseeps12}(b-c).  
Stable (upper) and unstable (lower) inverted states are seen to appear \emph{via} an isola-center bifurcation.
The edges of regression for elliptical hole inverted states can take more complicated multi-cusp forms not observed with circular holes.
Further details of these solutions can be found in Appendix \ref{se:ktauspecialpoints}.

In conclusion, we have examined the excision of high-energy material around an elastic singularity formed by inverting a simply folded thin disk. This process eliminates a source of rigidity, increasing the flexibility of the system. It reorients the low-curvature directions around the fold, influencing the opening angle distribution and eventually eliminating the inverted state, and thus the bistability, through a highly anisotropic mechanism re-entrant in the space of hole geometric parameters.
These findings have consequences for the mechanical compliance and energetics of perforated thin sheets, and for the design of deployable structures, in which fatigue of a highly stressed vertex is undesirable, prompting the introduction of gaps. Beyond folded and cut structures, similar mechanics is expected in other compliant mechanisms featuring networks of hinges, facets, and springs.
Bistability and critical hole dimensions are also influenced by the presence of multiple folds or the addition or removal of angular sectors of material to adjust the strength of the conical singularity, topics to be explored in detail in a future study \cite{yu2021bistability}.

\section*{Acknowledgments} 
TY and JH were partially supported by U.S. National Science Foundation grant CMMI-2001262.  MD thanks the Velux Foundations for support under the Villum Experiment program (Project No.\ 00023059).
We thank J. Gan for help with experiments, and M. Adda-Bedia and T. Jules for extensive discussions.  We also thank F. Lechenault, T. A. Witten, and J. Zhong for helpful discussions.

\clearpage

\appendix
\section{Numerical implementation of the developable model}\label{app:annulusmodel}

The inextensible strip model is formulated as a two-point boundary value problem (BVP), which can be parametrically studied using the continuation package AUTO 07P \cite{doedel2007auto}.
This requires normalizing the length of the integral interval to unity.  This additional step is not explicitly shown in the following discussion, but can be achieved by replacing the $s-$derivatives (primes) with derivatives with respect to a normalized parameter $s/s_\text{max}$ (if $s \in [0,s_\text{max}]$), thereby multiplying the right hand sides of all equations by $s_\text{max}$.  We retain the disk radius $R$ in the descriptions below, although in our calculations we set it equal to unity for simplicity.

To obtain inverted states, we begin with an annular sector of a flat disk subtending an angle $\rho < \pi$ and bend this into one half of a conical frustum, the initial solution for numerical continuation.  The angle $\rho$ is arbitrary and simply serves to create a nonflat conical starting point. Then the crease angle corresponding to a perfectly stiff crease is introduced by rotation of the ends of the sector, and $\rho$ is increased to $\pi$. The crease stiffness is relaxed by replacing the boundary constraints on the crease angle with conditions on the moment.  Finally, the hole dimensions are adjusted.

\begin{figure*}[h!]
	\centering
	\includegraphics[width=0.9\textwidth]{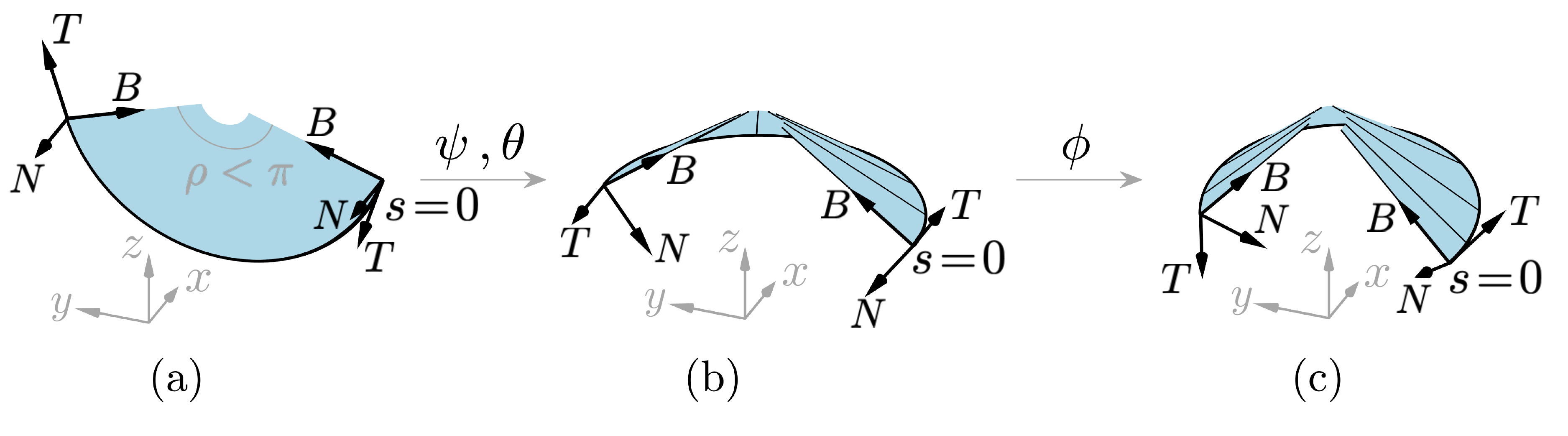}
	\caption{Euler angles are used to describe the sequential rotations of the director frame attached to the outer circle of an annular sector, following a $z$-$y$-$z$ (3-2-3) rotation convention. A Cartesian coordinate system is placed with cone axis $z$ and origin at $\tfrac{1}{2}\left(\bm{r} (0)+ \bm{r} (\rho R) \right)$. (a) The annular sector has its director frame attached to the outer circle; at the midpoint, $(\bm{-N},\bm{T},\bm{B})$ are aligned with $x,y,z$.  (b) The annular sector is deformed into a conical frustum.  The frame at each point is first aligned with the Cartesian axes in a manner akin to the midpoint, then rotated about $\bm{B}(s)$ by $\psi (s)$, and then about $\bm{T}(s)$ by $\theta (s)$. In this step, $\psi (s)$ is a linear function of $s$, and $\theta(s)$ is constant. (c) The crease angle is introduced by rotating the director frame at the two ends about $\bm{B}(0)$ and $\bm{B}(\rho R)$.}\label{appFig:Eulerrotation}
\end{figure*}

We employ Euler angles $(\psi, \theta, \phi)$ to relate the director frame to a fixed Cartesian frame \cite{love1927treatise},
\begin{widetext}
\begin{equation}\label{appeq:323rotation} 
\begin{bmatrix} \bm{-N} \\ \bm{T} \\ \bm{B} 
\end{bmatrix}
=
\begin{bmatrix} \cos \phi & \sin\phi & 0 \\ -\sin \phi & \cos \phi & 0 \\ 0 & 0 & 1 
\end{bmatrix}
\begin{bmatrix} \cos \theta & 0 & -\sin \theta \\ 0 & 1 & 0 \\ \sin \theta & 0 & \cos \theta 
\end{bmatrix}
\begin{bmatrix} \cos \psi & \sin\psi & 0 \\ -\sin \psi & \cos \psi & 0 \\ 0 & 0 & 1 
\end{bmatrix}
\begin{bmatrix} \uvc{x} \\ \uvc{y} \\ \uvc{z}
\end{bmatrix} \,.
\end{equation}
\end{widetext}
As $\kappa_n=-\bm{N}' \cdot \bm{T}\,, \kappa_g=\bm{B}' \cdot \bm{T} \,,$ and $ \tau_g=-\bm{B}' \cdot \bm{N}$, we have
\begin{align}\label{appeq:darbouxcurvature} 
\kappa_n=\phi' +\psi' \cos \theta  \,, \nonumber \\
 \kappa_g=  - \theta' \sin \phi + \psi' \sin \theta \cos \phi \,, \nonumber \\
 \tau_g =\theta' \cos \phi +\psi' \sin \theta \sin \phi \,.
\end{align}
Figure \ref{appFig:Eulerrotation} shows a Cartesian coordinate system $(x,y,z)$ and a sequence of rotations applied to deform the annular sector into one conical half of a creased structure with axis $z$ and bisected by the $x$-$z$ plane.  The origin is at $\tfrac{1}{2}\left(\bm{r} (0)+ \bm{r} (\rho R) \right)$.  

For circular holes, the limit $V$ of the generator coordinate $v$ can be obtained explicitly in terms of the function $\eta$ and the annular radii $a$ and $R$ \cite{dias2015wunderlich}. 
However, for elliptical holes, $V$ has a complicated dependence on the backbone coordinate $s$ and must be represented by an implicit function $\chi(V, s, \eta ;  a, b, R)=0$.  Using $\eta = \cot \beta $ and $\lambda=s/R$, this is
\begin{align}\label{appeq:elliptickai} 
	\chi(s, \eta, V(s, \eta)) = V^2 +(R^2-2V R -b^2) \sin^2 \beta \nonumber \\
	+ [(b/a)^2-1] \left[ V\cos(\beta + \lambda) + R\sin\beta\sin\lambda \right]^2 \, ,
\end{align}
which simplifies considerably for circular holes, for which $a=b$.
We treat $V$ as an independent variable and turn the algebraic constraint $\chi=0$ into a differential equation.
Denoting explicit partial derivatives by subscripts, we have $V'= V_s + V_{\eta} \eta'$
with $V_s = - \chi_s / \chi_V $ and $V_{\eta} = - \chi_{\eta} / \chi_V $ obtained through implicit differentiation \cite{starostin2015equilibrium}.
We further differentiate the algebraic constitutive law in \eqref{eq:stripgovern3} and combine with \eqref{eq:stripgovern4} to obtain a first order ordinary differential equation (ODE) for $\kappa_n$ and a second order ODE for $\eta$, and introduce another variable $\Omega \,\, (=\eta')$ to convert the latter to two first order ODEs.
The system is made autonomous by adding a trival differential equation $s'=1$.
Constants ``lost'' through differentiation are added back using additional boundary conditions.
All of this is combined with equations (\ref{eq:stripgovern1}-\ref{eq:stripgovern4}), \eqref{appeq:darbouxcurvature}, and $\bm{r}'(s)=\bm{T}$ to form the full system,
\begin{align}\label{appeq:17ODE} 
&F_1'-\kappa_n F_2 +\kappa_g F_3=0 \,, \nonumber \\
 &F_2'+\kappa_n F_1 -\kappa_n \eta F_3=0 \,, \nonumber \\
 &F_3'+\kappa_n \eta F_2 -\kappa_g F_1=0 \,,  \\
&M_1'-\kappa_n M_2 +\kappa_g M_3=0 \,, \nonumber \\
&M_2'+\kappa_n M_1-\kappa_n \eta M_3 -F_3=0 \,, \nonumber \\
&M_3'+\kappa_n \eta M_2 -\kappa_g M_1+F_2=0 \,,  \\
&\eta' = \Omega \,, \nonumber \\
&(AE-C^2) \Omega ' = (CB-AI) \Omega+ AG -CJ \,, \nonumber \\
&(AE-C^2) \kappa_n ' = (IC-BE) \Omega + JE -GC \,, \\
&V'=-\frac{\chi_s}{\chi_V}-\frac{\chi_{\eta}}{\chi_{V}} \Omega \,, \\
&\psi' =(\tau_g \sin \phi + \kappa_g \cos \phi)/ \sin \theta \,, \nonumber \\
&\theta ' =\tau_g \cos \phi  - \kappa_g \sin \phi \,, \nonumber \\
&\phi ' =\kappa_n  - (\kappa_g \cos \phi + \tau_g \sin \phi) / \tan \theta \,, \\
&x'=-\sin \psi \cos\phi -\cos\psi \sin \phi \cos\theta \,, \nonumber \\
&y'=\cos \psi \cos \phi -\sin \psi \sin \phi \cos \theta \,, \nonumber \\
&z'=\sin \theta \sin\phi \,, \\
&s'=1 \,,
\end{align}
in which 
\begin{align*}
A&=Y_{\kappa_n \kappa_n} W \,, \\
 B&= Y_{\kappa_n \eta} W + Y_{\kappa_n} W_{\eta} + Y_{\kappa_n} W_{V} V_{\eta} \,, \\
 C&= Y_{\kappa_n \eta'} W + Y_{\kappa_n} W_{\eta'} \,,\\
I&=Y_{\eta' \eta} W + Y_{\eta'} W_{\eta} +Y_{\eta'} W_{V} V_{\eta} \\
&+Y_{\eta}  W_{\eta'} +Y W_{\eta' \eta} +Y W_{\eta' V} V_{\eta} \,,\\
E&=Y_{\eta' \eta'} W + 2 Y_{\eta'} W_{\eta'} +W_{\eta' \eta'}Y \,, \\
 J&= \eta' M_1 -F_2 + \kappa_g (M_1 - \eta M_3) - Y_{\kappa_n} W_V V_s \,,\\
G&=Y_{\eta} W + Y W_{\eta} +Y W_{V} V_{\eta} \\
&- \kappa_n M_1  - Y_{\eta'} W_{V} V_{s} - Y W_{\eta' V} V_{s} \,,
\end{align*}

\begin{figure*}[h!]
	\centering
	\includegraphics[width=0.9\textwidth]{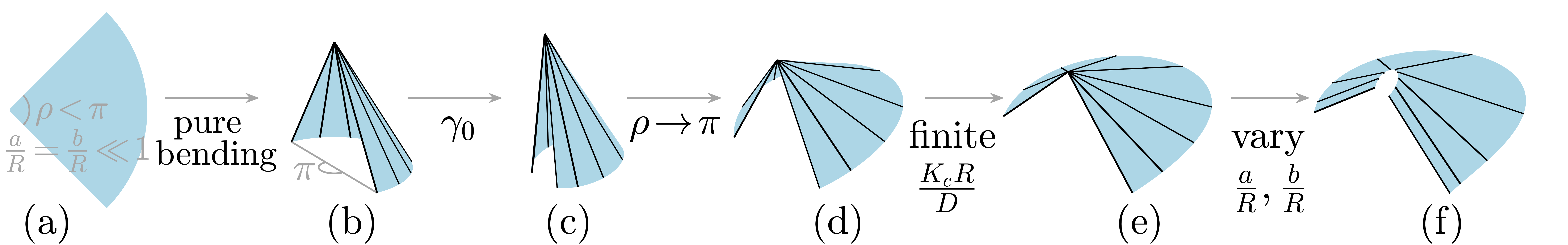}
	\caption{The inverted configuration is obtained through several continuation steps. (a) An annular sector that subtends an angle $\rho<\pi$ is (b) bent into a conical frustum with a central angle of $\pi$ and (c) has its ends rotated about themselves to introduce the crease angle $\gamma_0$.  Then (d) $\rho$ is increased to $\pi$, (e) a finite crease stiffness $K_cR/D$ is introduced, and (f) the hole dimensions $a/R$ and $b/R$ are varied.}
	\label{appFig:continuationstep}
\end{figure*}

Figure \ref{appFig:continuationstep} shows the sequence of continuation steps we use to obtain inverted solutions.  
A flat annular sector subtending an angle $\rho < \pi$ with a small circular hole $a \ll 1$ can be bent into the conical frustum of Figure \ref{appFig:continuationstep}b.  This starting solution for continuation is
\begin{align*}
&F_1=0\,, F_2=0\,, F_3=0\,, \\
&M_1=0\,, M_2=\ln \frac{a}{R} \,, M_3=- \sqrt{\frac{\pi^2}{\rho  ^2} - 1}\, \ln \frac{a}{R} \,, \\ 
&\kappa_n=\frac{1}{R}\sqrt{\frac{\pi^2}{\rho  ^2} - 1} \,, \eta=0\,, \eta '=0\,, \\
&V= R -a \, ,\\
&\psi= -\frac{\pi}{2} + \frac{\pi }{\rho R}s \,, \theta= -\sin^{-1} \frac{\rho}{\pi}\,,\phi=0\,, \\
&x=\frac{\rho R}{\pi } \sin \left (\frac{\pi}{\rho R}s \right)\,, y=-\frac{\rho R}{\pi } \cos \left (\frac{\pi }{\rho R}s \right) ,z=0 \, ,
\end{align*}
with $s \in \left[0 \, , \rho R \right]$.
The ends of the sector are rotated about themselves to introduce the rest crease angle $\gamma_0$ (Figure \ref{appFig:continuationstep}c).  The boundary conditions are then 
\begin{align*}
&F_z (0)=0\,, F_y (0)=0\,, \\
&M_x (0)=0\,,  \\
&\frac{\kappa_n (0) (1+\eta ^2 (0))^2}{\eta' (0) + \kappa_g (1+\eta^2 (0))} W(\eta' (0) , \eta (0)) \\
&\quad - \eta(0) M_1 (0) - M_3 (0)=0\,, \\
&\eta(0)=0\,, \eta(\rho R)=0\,, \\
&\chi(V(0),0,\eta(0))=0 \,,\\
&\psi(0)=-\frac{\pi}{2}\,, \phi (0)=-\left (\frac{\pi}{2} - \frac{\gamma_0}{2} \right)\,,\\
&\psi(\rho R)=\frac{\pi}{2}\,, \phi (\rho R)=\left (\frac{\pi}{2}  - \frac{\gamma_0}{2} \right)\,, \\
&x(0) =0 \,,z(0)=0,x(\rho R) =0\,, z(\rho R)=0\,, \\
&y(0)+y(\rho R)=0\,, \\ 
&s(0)=0\,,
\end{align*}
where symmetry dictates the conditions on $F_y(0)=\bm{F}(0) \cdot \uvc{y}$, $F_z(0)=\bm{F}(0) \cdot \uvc{z}$, and $M_x(0)=\bm{M}(0) \cdot \uvc{x}$.
At this point the boundary conditions constrain the crease angle, so we have a perfectly stiff crease.
In the next step, we increase $\rho$ to $\pi$ by a simple rescaling of $s/s_\text{max}$.
Then the actual crease stiffness is introduced by replacing the two boundary conditions for $\phi$ above with moment conditions
\begin{align*}
&M_3 (0) =K_{c} R/D (1-b/R) \sin \left[ \pi - \gamma_0 + 2 \phi (0) \right] \, , \\
 &M_3 (\pi R) =K_{c} R/D (1-b/R) \sin \left[ \pi - \gamma_0 - 2 \phi (\pi R{\tiny }) \right] \, , 
\end{align*}
and decreasing $K_{c} R/D$ from a large value to the real crease stiffness. 
Finally, the hole dimensions $a$ and $b$ are continued to their real values.

The geometry of the strip is reconstructed as 
\begin{widetext}
\begin{align}\label{appeq:3Dreconstruction} 
\bm{X}(s,v)&=\bm{r}(s)+v[\bm{B}(s)+\eta(s)\bm{T}(s)] \,,\\
&=(x-v [\eta (\sin \psi \cos \phi +\cos \psi \sin \phi \cos \theta)-\sin \theta \cos \psi]) \uvc{x}\nonumber \\
& +(y+v [\eta (\cos \psi \cos \phi -\sin \psi \sin \phi \cos \theta)+\sin \theta \sin \psi ]) \uvc{y}\nonumber \\
& +(z+v [\eta \sin \theta \sin \phi +\cos \theta ]) \uvc{z} \,,\nonumber
\end{align}
\end{widetext}
with $v \in [0,V]$. 

The edge of regression is 
\begin{align}\label{appeq:3Dedgeregression} 
\bm{c}(s)=\bm{r}(s)+ \frac{\sin \beta}{\beta'-\kappa_g}  \frac{\bm{B}(s)+\eta(s)\bm{T}(s)}{|\bm{B}(s)+\eta(s)\bm{T}(s)|} \nonumber \\
= \bm{r}(s)- \frac{\bm{B}(s)+\eta(s)\bm{T}(s)}{\eta'+\kappa_g(1+\eta^2)} \,.
\end{align}
Its first derivative is $\bm{c}'(s)=\tfrac{\eta \kappa_g^2  (1+\eta^2) + \eta''+3 \kappa_g \eta \eta' }{[\eta'+\kappa_g (1+\eta^2)]^2} (\bm{B} +\eta \bm{T})$.
At ``cylindrical'' points $\eta'=-\kappa_g (1+\eta^2)$, the edge of regression goes off to infinity, and the mean curvature is constant along the local generator. 
At ``conical'' points $\eta''=-3 \kappa_g \eta \eta' - \eta \kappa_g ^2 (1+\eta^2)$, the edge of regression has a cusp.
The generators can be mapped onto the flat annular sector with tangent $\bm{t}(s)$ and binormal $\bm{b}(s)$ using
\begin{align}\label{appeq:2Dreconstruction} 
\bm{X}(s,v)&=\bm{r}(s)+v[\bm{b}(s)+\eta(s)\bm{t}(s)] \,, \\
& =(R\sin \lambda - v \sin \lambda  + v \eta \cos \lambda )\uvc{x} \nonumber \\
&+ \left(- R \cos \lambda + v \eta \sin \lambda + v \cos \lambda \right) \uvc{y} \,,\nonumber
\end{align}
and the edge of regression onto it using
\begin{align}\label{appeq:2Dedgeregression}
\bm{c}(s)&=\bm{r}(s)- \frac{\bm{b}(s)+\eta(s)\bm{t}(s)}{\eta'+\kappa_g(1+\eta^2)} \,, \\
&= \left( R\sin \lambda +\frac{\sin \lambda  - \eta \cos \lambda}{\eta'+\kappa_g(1+\eta^2)} \right)\uvc{x} \nonumber\\
& - \left (R\cos \lambda + \frac{ \eta \sin \lambda \cos \lambda}{\eta'+\kappa_g(1+\eta^2)} \right) \uvc{y} \,.\nonumber
\end{align}

\section{Finite element simulations}\label{app:FEA}

\begin{figure*}[h]
	\centering
	\includegraphics[width=0.9\textwidth]{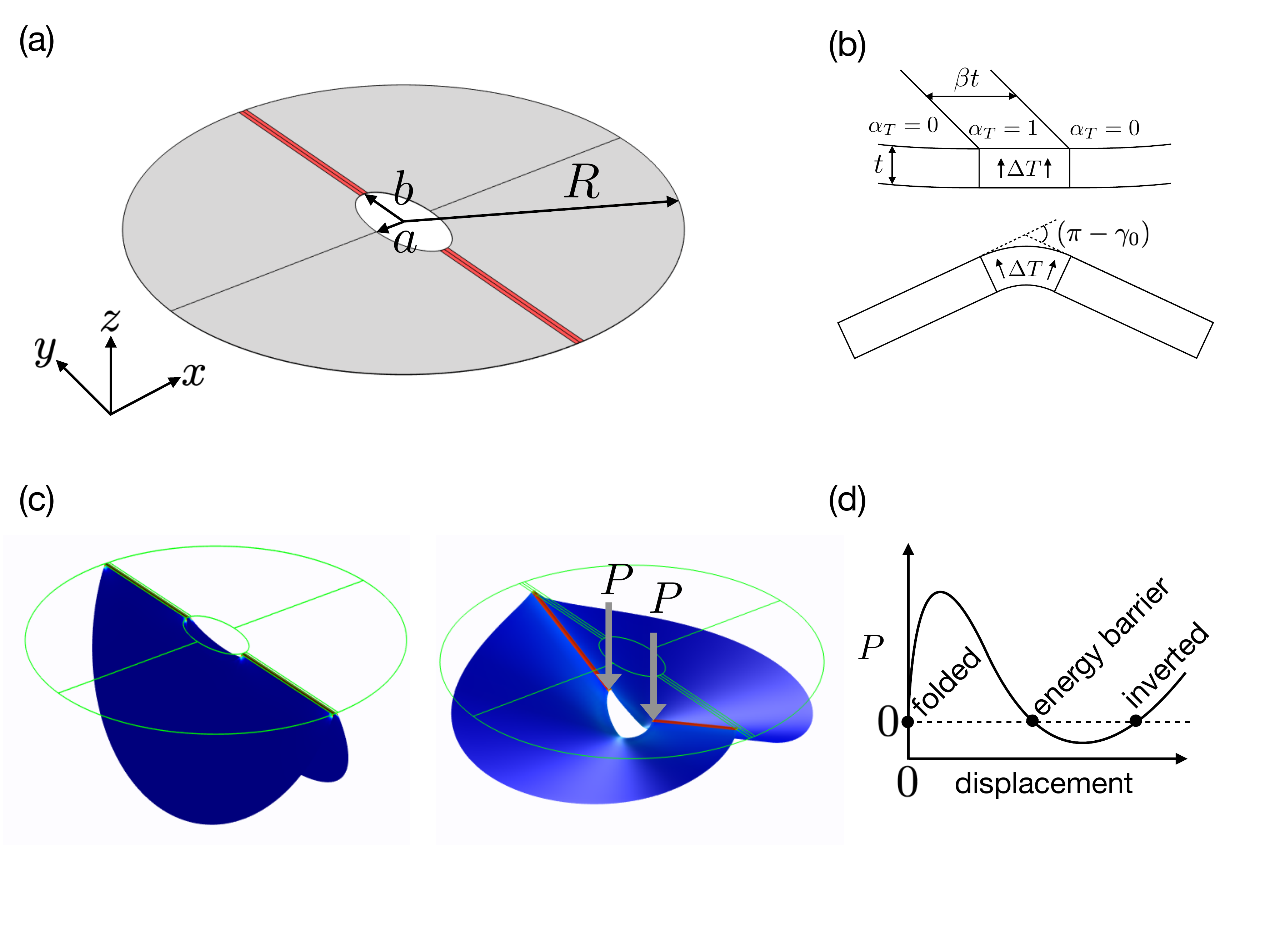}
	\caption{Aspects of FE simulations. (a) Geometry of the annulus. (b) Temperature-driven creasing. (c) Folded configuration and indentation process. (d) Schematic force-displacement curve. }
	\label{fig:scheme}
\end{figure*}

Simulations were performed in the commercial finite element (FE) software COMSOL Multiphysics 5.4.  We used quadratic shell elements with a linear elastic Hookean material and geometrically nonlinear kinematic relations, and searched for solutions with the default stationary solver that implements the nonlinear Newton method. Mesh refinement studies were undertaken to ensure convergence of the results.  Symmetries of the disk were exploited so that only one quarter of the domain required simulation.
Aspects of the simulations are illustrated schematically in Figure \ref{fig:scheme}.

The disk dimensions were set to $R=75 \,$mm and $t=127 \, \mu$m, and the material properties used were a Young's modulus $E=3.63$ GPa and Poisson's ratio $\nu=0.4$, consistent with reported material data and previous measurements on polyester sheets.

The simulations consist of two steps.  First, a temperature-induced folding angle is set via a coupled phase-field-like model, in which the effect of a through-thickness temperature gradient is confined to within the crease region. Then, with the two outer crease ends vertically constrained, a downward force is applied at both inner crease ends to deform the structure from the folded configuration to the inverted configuration.

The temperature-induced creasing follows the method introduced in \cite{andrade2019foldable}. 
An effective thermal expansion coefficient $\rho_T$ takes the value $1$ K$^{-1}$ in a strip of width $\beta t$ and $0$ elsewhere.  In the present work, we use $\beta=20$, so that the width of the crease is comparable with that of the plastic region of real creases in mylar sheets \cite{jules19local}.
The plate is subjected to a through-thickness temperature difference $\Delta T$ and responds with localized bending in the crease region.
Empirically, we have found that for folded rectangular sheets, the rest crease angle $\gamma_0$ is given approximately by
\begin{equation}
\gamma_0 = \pi - 1.4 \rho_T \Delta T \beta \, . \label{eq.FoldingAngle}
\end{equation}
By adjusting $\Delta T$ we are able to approximate a desired rest crease angle.  In the present work, while aiming for a $\gamma_0 = 45^\circ$ we achieved  $\gamma_0 = 48 \pm 1^\circ$ with most of the deviation occurring near the edges of the crease. 

During indentation, localized buckling can occur for some hole geometries.  To prevent this, foundation springs in the $x$ direction were attached to the crease ends at the beginning of indentation, whose stiffnesses decrease linearly with the vertical indentation depth so that their effects vanish before snapping occurs.

We estimate an effective crease stiffness for comparison with the developable model by ignoring boundary and ridge effects.  Let the moment per unit length of crease be
\begin{align}
M = \int^{t/2}_{t/2} z \sigma_{xx} dz\, ,
\end{align}
and the stress-strain relation be
\begin{align}
\sigma_{xx} = \frac{E}{1-\nu^2} (\epsilon_{xx}-\epsilon^{0}_{xx})\, ,
\end{align}
where $\epsilon^{0}_{xx}$ is the ``rest strain'' due to thermal expansion.  The strains are related to the crease angle by
\begin{align}
\epsilon_{xx} = \frac{\pi - \gamma }{\beta t} z\, .
\end{align}
Thus, the resultant moment on one half of the crease is
\begin{equation}
M = - (R-b) \frac{D}{\beta t} (\gamma - \gamma_0) \, .
\end{equation}
We can thus identify $\frac{D}{\beta t}$ with $K_c$ in the developable model, and $\frac{R}{\beta t}$ with the dimensionless crease stiffness $K_cR/D$, where $\beta = 20$ in the present work.

\section{Details of developable solutions}\label{se:ktauspecialpoints} 

\begin{figure*}[h!]
	\centering
	\includegraphics[width=\textwidth]{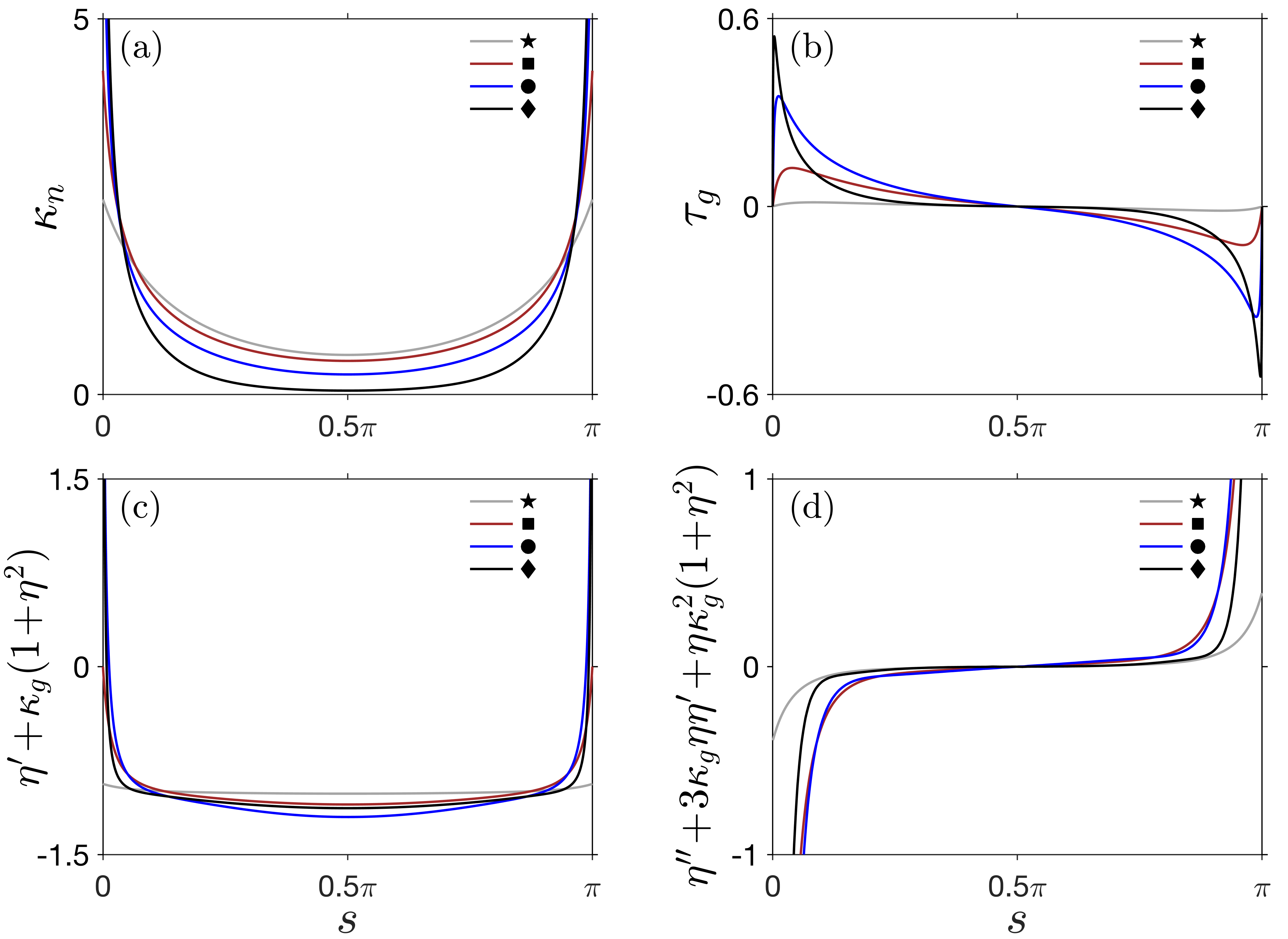}
	\caption{Details of the marked solutions in Figure \ref{fig:circularsolu}. (a) Normal curvature $\kappa_n$. (b) Geodesic torsion $\tau_g$. (c) Zeroes of $ {\eta^{\prime}} \!+\! \kappa_g(1 \!+\! \eta^2) $ correspond to cylindrical points, where the edge of regression goes off to infinity. (d) Zeroes of $ \eta^{\prime \prime} \!+\! 3 \kappa_g \eta {\eta^{\prime}} \!+\! \eta  \kappa_g^2 (1 \!+\! \eta^2) $ correspond to conical points, where the edge of regression has cusps.}\label{appfig:Circularktcusp}
\end{figure*}

\begin{figure*}[h!]
	\centering
	\includegraphics[width=\textwidth]{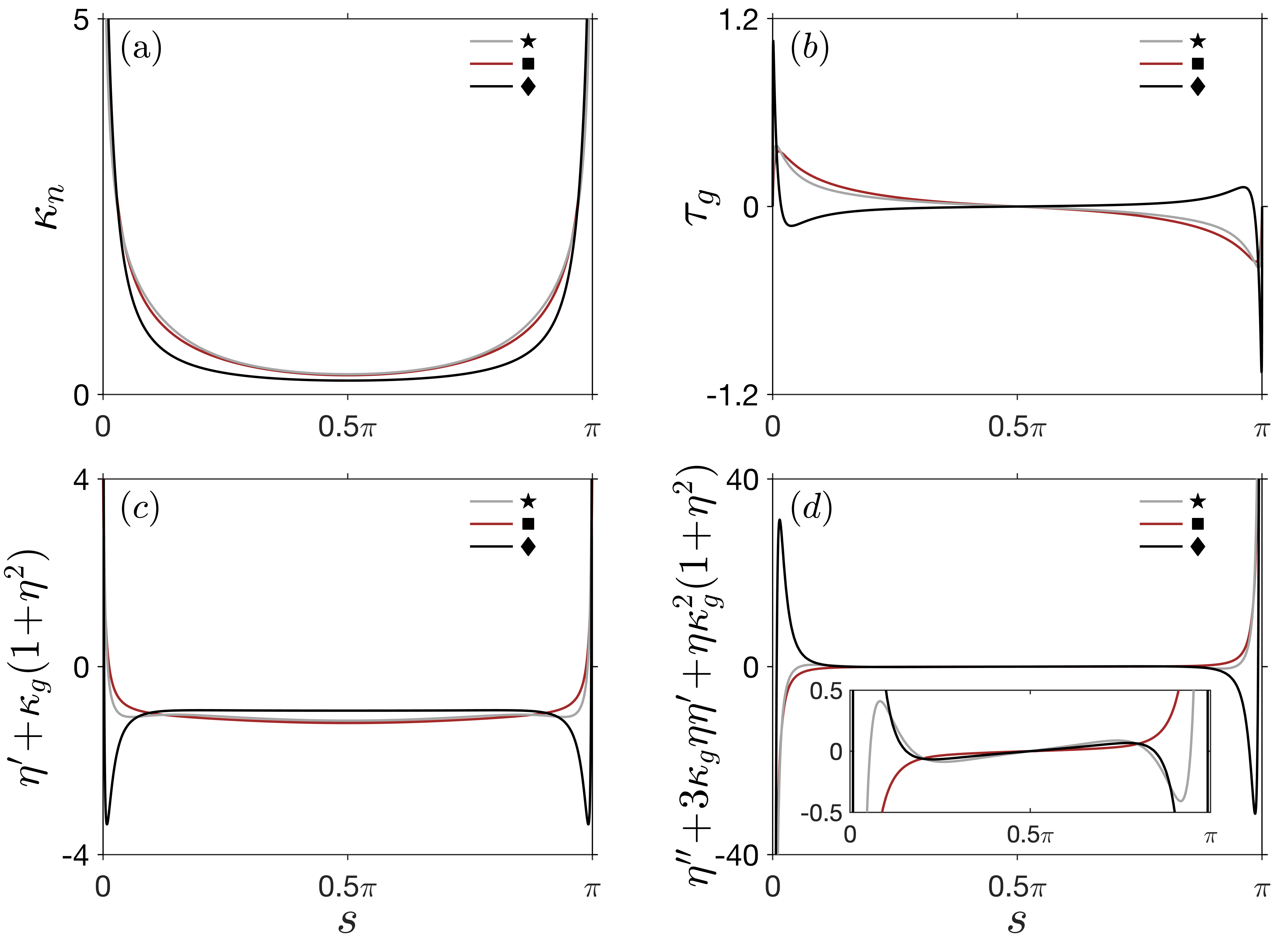}
	\caption{Details of the marked solutions in Figure \ref{fig:phaseeps12}. (a) Normal curvature $\kappa_n$. (b) Geodesic torsion $\tau_g$. (c) Zeroes of $ {\eta^{\prime}} \!+\! \kappa_g(1 \!+\! \eta^2) $ correspond to cylindrical points, where the edge of regression goes off to infinity. (d) Zeroes of $ \eta^{\prime \prime} \!+\! 3 \kappa_g \eta {\eta^{\prime}} \!+\! \eta  \kappa_g^2 (1 \!+\! \eta^2) $ correspond to conical points, where the edge of regression has cusps.}\label{appfig:ellipticktcusp}
\end{figure*}

Figures \ref{appfig:Circularktcusp} and \ref{appfig:ellipticktcusp} provide details of the geometry of configurations from Figures \ref{fig:circularsolu} and \ref{fig:phaseeps12}, respectively.

\clearpage

\bibliographystyle{unsrt}

\end{document}